\documentclass[conference]{IEEEtran}
\IEEEoverridecommandlockouts

\usepackage[dvipsnames]{xcolor}
\usepackage{color}

\definecolor{PPQAblue}{HTML}{045275} 
\definecolor{PPQAgreen}{HTML}{24B194}
\definecolor{PPQAlightgreen}{HTML}{94CEB9}
\definecolor{PPQAorange}{HTML}{EB5A36}
\definecolor{PPQApink}{HTML}{DC3977} 
\definecolor{PPQApurple}{HTML}{7C1D6F}

\ifCLASSOPTIONcompsoc
    \usepackage[caption=false, font=normalsize, labelfont=sf, textfont=sf]{subfig}
\else
\usepackage[caption=false, font=footnotesize]{subfig}
\fi

\usepackage{adjustbox}
\usepackage[utf8]{inputenc}
\usepackage{physics}% The copyright text:

\makeatletter
\def\ps@IEEEtitlepagestyle{%
  \def\@oddfoot{\mycopyrightnotice}%
  \def\@evenfoot{}%
}
\def\mycopyrightnotice{%
  {\footnotesize
  \hfill 
  \parbox{\textwidth}{%
  © 2025 IEEE.  Personal use of this material is permitted.  Permission from IEEE must be obtained for all other uses, in any current or future media, including reprinting/republishing this material for advertising or promotional purposes, creating new collective works, for resale or redistribution to servers or lists, or reuse of any copyrighted component of this work in other works.}
  \hfill}
}
\makeatother
\usepackage{bm} % bold math
\usepackage{bbm} % black board math
\usepackage{braket}
\usepackage{dsfont}
\usepackage{amsmath}
\usepackage{amssymb}
\usepackage{amsthm}
\usepackage{thmtools} % for correct autoref naming
\usepackage{mathtools}

\makeatletter
\newcommand{\gettikzxy}[3]{%
  \tikz@scan@one@point\pgfutil@firstofone#1\relax
  \edef#2{\the\pgf@x}%
  \edef#3{\the\pgf@y}%
}
\makeatother
\usepackage{graphics}
\usepackage{graphicx}
\usepackage{float}

\usepackage[colorlinks]{hyperref}
\hypersetup{
    colorlinks  = true,
    citecolor   = RoyalPurple,
    linkcolor   = teal,
    urlcolor    = teal
}

\usepackage{etoolbox}
\apptocmd{\sloppy}{\hbadness 10000\relax}{}{}

\usepackage{silence}

        \DeclareMathOperator{\QKP}{QKP} % quadratic knapsack problem
        \DeclareMathOperator{\MDKP}{MDKP} % multi-dimensonal knapsack problem
        \DeclareMathOperator*{\ancilla}{a} % ancilla
        \newcommand*{\C}{\mathbb{C}}
        \newcommand*{\hil}{\mathcal{H}}
        \newcommand*{\hamming}{\Delta} % Hamming distance
        \newcommand*{\numbits}[1]{\abs{{#1}}_{\text{bin}}} % length of binary representation
        \newcommand*{\qkp}{\QKP_{n}(\bm{p}, \bm{w}; c)} % generic quadratic knapsack instance
        \newcommand*{\mdkp}{\MDKP_{n}(\bm{p}, \bm{w}; \bm{c})} % generic multi-dimensional knapsack instance
        \newcommand*{\QTG}{\mathcal{G}} % state preparation circuit
        \newcommand*{\controlled}{C} % control
\usepackage{tcolorbox}  % <---
\usepackage{ragged2e}
\usepackage{enumitem}

\makeatletter
\def\ps@IEEEtitlepagestyle{%
  \def\@oddfoot{\mycopyrightnotice}%
  \def\@evenfoot{}%
}
\def\mycopyrightnotice{%
  {\footnotesize
  \hfill 
  \parbox{\textwidth}{%
  © 2025 IEEE.  Personal use of this material is permitted.  Permission from IEEE must be obtained for all other uses, in any current or future media, including reprinting/republishing this material for advertising or promotional purposes, creating new collective works, for resale or redistribution to servers or lists, or reuse of any copyrighted component of this work in other works.}
  \hfill}
}
\makeatother

\begin{document}

\title{A quantum search method for quadratic and multidimensional knapsack problems}
\author{
\IEEEauthorblockN{
    S\"oren Wilkening\IEEEauthorrefmark{1}${}^{,}$\IEEEauthorrefmark{2}, 
    Andreea-Iulia Lefterovici\IEEEauthorrefmark{1}, 
    Lennart Binkowski\IEEEauthorrefmark{1},\\
    Marlene Funck\IEEEauthorrefmark{1},
    Michael Perk\IEEEauthorrefmark{3},
    Robert Karimov\IEEEauthorrefmark{1},
    Sándor Fekete\IEEEauthorrefmark{3},
    Tobias J. Osborne\IEEEauthorrefmark{1}
}
\IEEEauthorblockA{\IEEEauthorrefmark{1}Institut f\"ur Theoretische Physik, Leibniz Universit\"at Hannover, Hannover, Germany}
\IEEEauthorblockA{\IEEEauthorrefmark{2}Email: soeren.wilkening@itp.uni-hannover.de}
\IEEEauthorblockA{\IEEEauthorrefmark{3}Deparment of Computer Science, Technische Universit\"at Braunschweig, Braunschweig, Germany}
}
\maketitle

% adds page numbers
%\thispagestyle{plain}
%\pagestyle{plain}

\begin{abstract}
    Solving combinatorial optimization problems is a promising application area for quantum algorithms in real-world scenarios.
    In this work, we extend the ``Quantum Tree Generator'' (QTG), previously proposed for the $\bm{0}$-$\bm{1}$ Knapsack Problem, to the $\bm{0}$-$\bm{1}$ Quadratic Knapsack Problem (QKP) and the Multidimensional Knapsack Problem (MDKP).
    The QTG constructs a superposition of all feasible solutions for a given instance and can therefore be utilized as a promising state preparation routine within amplitude amplification to produce high-quality solutions.
    Previously, QTG-based search was tested on the $\bm{0}$-$\bm{1}$ Knapsack Problem, where it demonstrated the potential for practical quantum advantage, once quantum computers with a few hundred logical and fully connected qubits are available.
    Here, we evaluate the algorithm's performance on QKP and MDKP against the classical solver Gurobi.
    To facilitate large-scale evaluations, we employ an advanced benchmarking technique that enables runtime predictions for instances with up to $\bm{2000}$ variables for QKP and up to $\bm{1500}$ variables and $\bm{100}$ constraints for MDKP.
    Our results indicate that QTG-based search can produce high-quality solutions with competitive runtimes for QKP. 
    However, its performance declines for MDKP, highlighting the challenges quantum algorithms face when tackling highly constrained optimization problems.
\end{abstract}


\section{\label{section:Introduction}Introduction}

In recent years, the potential of real-world quantum computing has fueled optimism about solving complex combinatorial optimization problems, prompting extensive theoretical research into developing and analyzing effective quantum methods.
% And in the coming years, quantum information processing devices are expected to incorporate many thousands of physical qubits~\cite{ResearchIBM2023Charting100000Qubits}.
So far, the development of new quantum algorithms has largely been driven by theoretical measures, especially worst-case asymptotic runtime analysis~\cite{Brassard2002QuantumAmplitudeAmplificationAndEstimation,Harrow2009QuantumAlgorithmForLinearSystemsOfEquations,Montanaro2020QuantumSpeedupOfBranchAndBoundAlgorithms}.
However, a major challenge remains in demonstrating quantum advantage for \emph{practical applications}.
Recent work targeted combinatorial optimization problems that provide a rich set of application areas~\cite{Ajagekar2020QuantumComputingBasedHybridSolutionStrategiesForLargeScaleDiscreteContinuousOptimizationProblems,Wilkening2023AQuantumAlgorithmForTheSolutionOfTheKnapsackProblem}, but also have strong classical competitors:
For these problems, classical solvers like Gurobi~\cite{Gurobi}, CPLEX~\cite{CPLEX} or CP-SAT~\cite{CPSAT} routinely solve problems involving thousands of variables to \emph{provable optimality}.
An emerging benchmarking discipline~\cite{Cade2023QuantifyingGroverSpeedUpsBeyondAsymptoticAnalysis,Ammann2023RealisticRuntimeAnalysisForQuantumSimplexComputation,Wilkening2023AQuantumAlgorithmForTheSolutionOfTheKnapsackProblem,Lefterovici2025BeyondAsymptoticScalingComparingFunctionalQuantumLinearSolvers} is to combine classical emulation techniques, quantitative complexity bounds, and sophisticated, problem- and algorithm-specific simulations to benchmark quantum algorithm against the classical powerhouses for problem instances far larger than the limit of roughly 50 qubits imposed by state vector-based simulation~\cite{Lykov2023FastSimulationOfHighDepthQAOACircuits}.

The novel Quantum Tree Generator (QTG) search method proposed by
Wilkening~\textit{et~al.}~\cite{Wilkening2023AQuantumAlgorithmForTheSolutionOfTheKnapsackProblem}
combines classical structural insights with 
properties of quantum algorithms to first generate all feasible solutions in superposition, from which a high-quality solution is extracted with quantum 
amplitude amplification~\cite{Brassard2002QuantumAmplitudeAmplificationAndEstimation} and showed potential \emph{practical} quantum advantage for mid-scale instances of the $0$-$1$ Knapsack Problem (KP).
It is worth noting that QTG comes with some theoretical performance guarantees at the expense of an exponential worst-case runtime.
Running the quantum maximum finding routine (QMaxSearch) for $\leq \sqrt{2^{n}}$ iterations makes QTG similar to an exact algorithm, as it finds
an optimal solution with probability $\frac{1}{2}$~\cite{Durr1996AQuantumAlgorithmForFindingTheMinimum}. 
However, in practice, we limit the maximum number of Grover iterations to a polynomial $M$ (as discussed in \autoref{section:QuadraticKnapsack} and \autoref{section:MultidimensionalKnapsack}), which ensures polynomial runtime at the expense of theoretical performance guarantees, while still finding high-quality solutions with high probability in practice.

In this work, we utilize and adapt the QTG to solve realistic instances of the $0$-$1$ Quadratic Knapsack (QKP) and the $0$-$1$ Multidimensional Knapsack Problem (MDKP).
We then compare QTG with results obtained by the integer programming and quadratic programming solver Gurobi~\cite{Gurobi}.
As the QTG ultimately is neither a tailored algorithm for QKP nor MDKP the comparison with a general-purpose solver is both fair and interesting.
To evaluate the performance of the QTG-based search on realistic benchmark instances, we infer its expected number of cycles. 
Under some benevolent assumptions for quantum hardware (see \autoref{subsection:AssumptionsOnTheQuantumProcessingUnit}) our results indicate that the QTG-based method can require less time to find high-quality solutions, and in the best case, the optimal solution for realistic instances already at $100$ variables.
To ease the reader into the paper, we use the notation provided by \cite{Wilkening2023AQuantumAlgorithmForTheSolutionOfTheKnapsackProblem} that is summarized in \autoref{table:Notation}.

\subsection{\label{subsection:AssumptionsOnTheQuantumProcessingUnit}Assumptions on the Quantum Processing Unit}

The resource estimation setup for applying the QTG method to QKP and MDKP relies on several fault-tolerant assumptions for the quantum processing unit (QPU).
It assumes that all qubits and gates are noiseless, logical components. 
The universal gate set includes all single-qubit gates, singly-controlled single-qubit rotations, and Toffoli gates, all of which have equal implementation cost. 
Disjoint gates can be executed in parallel, counting as a single cycle on the QPU. 
Singly-controlled single-qubit rotations and Toffoli gates can be applied to arbitrary tuples of qubits.
Additionally, all multi-controlled gates have access to the same ancilla qubits and implementing a control on $0$ is just as costly as on $1$.
Finally, to calculate a runtime from the cycle counts, we assume a quantum cycle time limit of $1ns$ given by \cite{Chew2022UltrafastEnergyExchangeBetweenTwoSingleRydbergAtomsOnANanosecondTimescale}.

\begin{table}
    \begin{center}
    \bgroup
\def\arraystretch{1.5}
\begin{tabular}{ c | c }
 Notation & Interpretation\\ 
 \hline
$\hil_{1} \ni \ket{\bm{x}}^{1}$ & n-qubit path register \\ 
$\hil_{2} \ni \ket{c_{\bm{x}}}^{2}$ & capacity register \\ 
$\hil_{3} \ni \ket{P_{\bm{x}}}^{3}$ & profit register \\
$\hil_{a} \ni \ket{0}^{a} $ & ancilla register\\ 
$\QTG =  \prod_{m = 1}^{n} U_{m}$ & QTG circuit; composed of n-layer unitaries \\
$U_{m} = U_{m}^{3} U_{m}^{2} U_{m}^{1}$ & components of each layer unitary\\
$ U^{1}_{m} = \controlled^{2}_{\geq w_{m}}  \left(R_{y,b}^{x_m}\right)$ & branching unitary\\
$U^{2}_{m} = \controlled^{1}_{m}\left( \text{SUB}_{w_{m}} \right)$ & QFT subtraction unitary\\
$U^{3}_{m} = \controlled^{1}_{m}\left( \text{ADD}_{p_{m}} \right)$ & QFT addition unitary\\
$R_{y,b}^{x_m}$ & $R_y$ gate with a bias $b$ \\\vspace{-0.17cm}
$C_{\geq w_{m}}^{2}$ & control on whether the 2nd register\\ 
 &state is greater or equal to the integer $w_{m}$\\\vspace{-0.17cm}
$\controlled^{1}_{m}$ & subtraction and addition are controlled \\
&on the $m$-th qubit in the first register\\
\hline
\end{tabular}
\egroup
\end{center}
\caption{\label{table:Notation}Notation for the following sections from \cite{Wilkening2023AQuantumAlgorithmForTheSolutionOfTheKnapsackProblem}.}
\end{table}

\subsection{Related Work}

\label{sec_RelatedWrok}

The QKP was first introduced by Gallo~\textit{et~al.}~\cite{Gallo1980QuadraticKnapsackProblems} and has a straightforward quadratic programming formulation (see~\autoref{section:QuadraticKnapsack}).
Its practical solvability often depends on instance properties such as density~\cite{Pisinger2007TheQuadraticKnapsackProblemASurvey,Cacchiani2022KnapsackProblemsAnOverviewOfRecentAdvancesPartMultipleMultidimensionalAndAuadraticKnapsackProblems}.
This has drawn significant interest from the optimization community, leading to recent exact algorithms like those in~\cite{DjeumouFomeni2022ACutAndBranchAlgorithmForTheQuadraticKnapsackProblem,Hochbaum2025AFastAndEffectiveBreakpointsHeuristicAlgorithmForTheQuadraticKnapsackProblem} as well as heuristic approaches such as~\cite{FENNICH2024102,DJEUMOUFOMENI202352}.
Other approaches include a quantum-inspired evolutionary algorithms~\cite{Narayan2009ANovelQuantumEvolutionaryAlgorithmForQuadraticKnapsackProblem}.
In contrast, the MDKP tends to be more tractable for integer programming solvers such as Gurobi~\cite{Vimont2008ReducedCostsPropagationInAnEfficientImplicitEnumerationForTheMultidimensionalKnapsackProblem}.
Recent approaches, such as those in~\cite{Setzer2020EmpiricalOrthogonalConstraintGenerationForMultidimensionalKnapsackProblems}, have proven particularly effective for solving large instances~\cite{Cacchiani2022KnapsackProblemsAnOverviewOfRecentAdvancesPartMultipleMultidimensionalAndAuadraticKnapsackProblems}.
As for the QKP, heuristic approaches~\cite{KONG20157,ANGELELLI20102017,LAI2018282} as well as quantum-inspired heuristics \cite{Haddar2016AHybridQuantumParticleSwarmOptimizationForTheMultidimensionalKnapsackProblem} have been studied.

Apart from QTG, many other quantum algorithms and frameworks for solving combinatorial optimization problems have been proposed and studied, such as \emph{Quantum Annealing} \cite{Kadowaki1998QuantumAnnealingInTheTransverseIsingModel}, \cite{Mohseni2022IsingMachinesAsHardwareSolversOfCombinatorialOptimizationProblems}, \cite{ Lo2023AnIsingSolverChipBasedOnCoupledRingOschillators}, \cite{Pusey2020AdiabaticQuantumOptimizationFailsToSolveKnapsackProblem}, \emph{Nested Quantum Search} (NQS) \cite{Cerf2000NestedQuantumSearchAndStructuredProbleems}, \emph{Quantum Approximate Optimization Algorithm} (QAOA) \cite{Farhi2014AQuantumApproximateOptimizationAlgorithm} and it's generalization to the \emph{Quantum Alternating Operator Ansatz} \cite{Hadfield2019FromTheQuantumApproximateOptimizationAlgorithmToAQuantumAlternatingOperatorAnsatz}, \emph{Quantum Backtracking} \cite{Montanaro2018QuantumWalkSpeedupOfBacktrackingAlgorithms}, \cite{Martiel2020PracticalImplementationQuantumBacktrackingAlgorithm}, and \emph{Quantum Branch-and-Bound} (QBnB) \cite{Montanaro2020QuantumSpeedupOfBranchAndBoundAlgorithms}.
All of these frameworks are, in principle, also applicable to a variety of knapsack problems.
For example, while NQS and QBnB theoretically promise an advantage over classical solvers, Wilkening~\textit{et~al.}~\cite{Wilkening2023AQuantumAlgorithmForTheSolutionOfTheKnapsackProblem} showed they are significantly outperformed by both the QTG search method and classical solvers on large real-world benchmarks when applied to the KP.

Several variants of QAOA have been to the ordinary KP such as in \cite{VanDam2021QuantumOptimizationHeuristicsWithAnApplicationToKnapsackProblems} (introduces a hyperparameter-free objective function and special copula mixers), \cite{Christiansen2024QuantumTreeGeneratorImprovesQAOAStateOfTheArtForTheKnapsackProblem} (combines the QTG with the Grover-mixer QAOA framework \cite{Baertschi2020GroverMixersForQAOAShiftingComplexityFromMixerDesignToStatePreparation}), and \cite{Bucher2025IFQAOAAPenaltyFreeApproachToAcceleratingConstrainedQuantumOptimization} (combines QAOA with quantum phase estimation techniques \cite{Kitaev1995QuantumMeasurementsAndTheAbelianStabilizerProblem}).
Recently, the scope of QAOA has also been enlarged to the QKP \cite{Ha2024SolvingQuadraticKnapsackProblemWithBiasedQuantumStateOptimizationAlgorithm} and the MDKP \cite{Guney2025QUBOFormulationsAndCharacterizationOfPenaltyParametersForTheMultiKnapsackProblem}.

Another established quantum approach to the QKP is a quantum walk-based genetic algorithm \cite{Pizchai2015QuantumWalkBasedGeneticAlgorithmForQuadraticKnapsackProblem} while \cite{Cui2024HybridQuantumSearchAlgorithmForSolvingTheMultiDimensionalKnapsackProblem} applied the framework of quantum maximum finding \cite{Durr1996AQuantumAlgorithmForFindingTheMinimum} to the MDKP.
The latter therefore utilizes the same framework as our approach, but does not incorporate a customized problem-specific state preparation and diffusion circuit like the QTG.

\section{\label{section:QuadraticKnapsack}Quadratic Knapsack}

\begin{figure}
    \begin{tcolorbox}[
        colback=white, 
        colframe=teal!40,
        coltitle=black, 
        title=\noindent\justifying{Box 1: QTG for Quadratic Knapsack}]
    \begin{minipage}[t]{0.9\textwidth}
        \justifying\footnotesize
        \noindent\\\textbf{Input}\\
        Quadratic Knapsack instance $\qkp$.\\
        \\
        \textbf{1. Initialize:}\\
        Initialize the registers in the state $\ket{\bm{0}}^{1} \ket{c}^{2} \ket{0}^{3}$.\\\\
        \textbf{2. Tree traversing:}\\
        For each item $m = 1, \ldots, n$:\\
        \textbf{(a)} Create a superposition: $U^{1}_{m} = \controlled^{2}_{\geq w_m}  \left(R_{y,b}^{x_m}\right)$\\\\
        \textbf{(b)} Update the remaining costs: $U^{2}_{m} = \controlled^{1}_{m}\left( \text{SUB}_{w_{m}} \right)$\\\\
        \textbf{(c)} Update the total profit: $U^{3}_{m} = \controlled^{1}_{m}\left( \text{ADD}_{p_{m}} \right)$ \\

        \hspace{2.7cm} $U^{3}_{m m'} =  C_{m  m'}^{1}\left(\text{ADD}_{p_{m m'}}\right)$.
    \end{minipage}
    \label{box:QTG-qkp}
    \end{tcolorbox}
\end{figure}

Given is a set of $n$ items where any given item $m$ has an integer weight $w_{m}$. 
Additionally, we are given an $n \times n$ integer matrix $P = (p_{m m'})$, where $p_{m m}$ is the profit achieved if item $m$ is selected and $p_{m m'} + p_{m' m}$ is the profit achieved if both items $m$ and $m'$ are selected for $m < m'$. 
The QKP asks for a subset of items of maximum profit, such that their cumulative weights do not exceed some capacity $c$.
Without loss of generality, we assume that $\max_{m} w_{m} \leq c < \sum_{m} w_{m}$ and that the profit matrix is symmetric, i.e., $p_{m m'} = p_{m' m}$.
The QKP and some variations are known to be NP-hard~\cite{Pisinger2007TheQuadraticKnapsackProblemASurvey}.
In the following, we address a given instance of the QKP as $\qkp$.
It can be formulated as an integer quadratic program (IQP) as follows:
\begin{align*}
\begin{split}
    \text{maximize } & \sum_{m = 1}^{n} p_{m} x_{m} + \sum_{m = 1}^{n}\sum_{m'\neq m}^{n} p_{m m'} x_{m} x_{m'}\\
    \text{subject to } & \sum_{m = 1}^{n} w_{m} x_{m} \leq c \\
    & x_{m} \in\{0, 1\}, \quad m = 1, \ldots, n.
\end{split}
\end{align*}
The binary variable $x_{m}$ encodes whether item $m$ is selected.
A complete assignment of all items (also called \emph{path}) constitutes a bit string of length $n$.

We follow the quantum encoding strategy \cite{Wilkening2023AQuantumAlgorithmForTheSolutionOfTheKnapsackProblem} of representing the paths $\bm{x} = x_{1} \cdots x_{n}$ as computational basis states $\ket{\bm{x}}^{1}$ in an $n$-qubit \emph{path register} $\hil_{1} = \C^{2^{n}}$.
We store the remaining capacity $c - \sum_{m = 1}^{n} w_{m} x_{m}$ as a quantum state $\ket{c_{\bm{x}}}^{2}$ in a $\numbits{c}$-qubit \emph{capacity register} $\hil_{2}$ and record the total profit as a state $\ket{P_{\bm{x}}}^{3}$ in a $\numbits{P}$-qubit \emph{profit register} $\hil_{3}$, where $P$ is some upper bound on the optimal profit.
Additionally, we optimize the depth of the circuit at the expense of using $\max(n, \numbits{c}, \numbits{P})$ ancilla registers.
This leads to the total number of qubits in the composite register $\hil = \hil_{1} \otimes \hil_{2} \otimes \hil_{3} \otimes \hil_{\ancilla}$ of being
\begin{equation*}
%\label{equation:TotalNumberOfQubits} removed as its not used.
    n + \numbits{c} + \numbits{P} + \max(n, \numbits{c}, \numbits{P}).
\end{equation*}

\subsection*{Methods}

The QTG $\QTG$ for the QKP follows the same design criteria as its counterpart for the ordinary $0$-$1$ Knapsack Problem~\cite{Wilkening2023AQuantumAlgorithmForTheSolutionOfTheKnapsackProblem}.
We give here an independent introduction of the QTG, but draw comparisons to it's original formulation whenever suitable.
The entire state preparation part contains $n$ \emph{layer unitaries} $U_{m} = U_{m}^{3} U_{m}^{2} U_{m}^{1}$ which correspond to including/excluding each item in superposition while keeping track of the remaining capacity and the total profit arising from both the diagonal profits $p_{m}$ and the off-diagonal profits $p_{m m'}$.
Throughout this process, feasibility is ensured by controlling the branching at each item $m$ on the comparison ``$w_{m} \leq c_{\bm{x}}$''.
That is, we create a superposition of item $m$ being excluded/included if and only if its weight does not surpass the remaining capacity stored in the second register $\hil_{2}$.
Correspondingly, if the latter expression is not valid the $m$-th qubit stays in the $\ket{0}$-state, indicating the item's exclusion.
The corresponding unitary is denoted $U_{m}^{1} = C_{\geq w_{m}}^{2}(R_{y,b}^{x_m})$, where the $R_y$ gate is responsible for creating a biased superposition of item exclusion ($\ket{0}$-state) and inclusion ($\ket{1}$-state)
\begin{align*}
%\label{equation:Rygate} removed as its not used.
    R_{y,b}^{x_{m}}\hspace*{-3pt} =\hspace*{-3pt} \frac{1}{\sqrt{b+2}}\begin{pmatrix} \sqrt{1 + (1-x_{m})b} &\hspace*{-9pt} \sqrt{1 + x_{m} b} \\ \sqrt{1 + x_{m} b} &\hspace*{-9pt} -\sqrt{1 + (1-x_{m})b}\end{pmatrix}.
\end{align*}
For $b = 0$, we recover the Hadamard gate, mapping $\ket{0}$ to a uniform superposition of $\ket{0}$ and $\ket{1}$.
This corresponds to including or excluding the items with equal probability.
A non-zero bias $b$ helps maximizing the sampling probability of paths $\bm{x}$ with specific Hamming distance $\hamming \coloneqq \hamming(\bm{x}',\bm{x})$ to a given incumbent $\bm{x}'$.
In \cite{Wilkening2023AQuantumAlgorithmForTheSolutionOfTheKnapsackProblem} it is shown that the optimal choice for $b$ for fixed $\hamming$ is $b_{\text{opt}} \approx n / \hamming$.
Hence, for $b \to \infty$, we will recover the incumbent with certainty (as $\Delta \to 0$) while finite values allow us to target specific neighborhoods.
Our numerical results indicate that fine-tuning the bias to target distance-4 solutions significantly increases the probability of sampling improved states via the QTG.

\begin{figure}[H]
    \centering
    \includegraphics[width=0.8\linewidth]{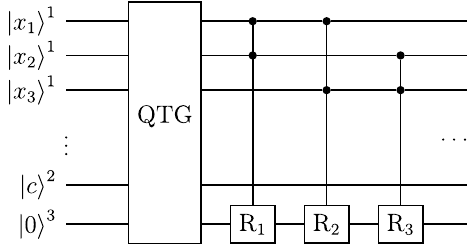}
    \caption{The circuit for QKP builds on the circuit QTG for KP \cite{Wilkening2023AQuantumAlgorithmForTheSolutionOfTheKnapsackProblem}, and takes into account the two-control-one-target gates for adding the quadratic profits on the profit register.}
    \label{figure:qknap_circuit}
\end{figure}

With the possible inclusion of item $m$ we have to update the remaining capacity as well as the total profit -- both accounting for the linear profit $p_{m}$ as well as all potentially gained quadratic profits $p_{m m'}$.
Updating the capacity amounts to subtracting $w_{m}$ from the capacity register's state, controlled on the $m$-th qubit: $U_{m}^{2} = C_{m}^{1}(\text{SUB}_{w_{m}})$, realized e.g. via QFT-based addition circuits \cite{Draper2000AdditionOnAQuantumComputer}.
Analogously, accounting for the linear profit translates into a controlled addition of $p_{m}$ in the profit register $C_{m}^{1}(\text{ADD}_{p_{m}})$.
Furthermore -- and here lies the main difference between the QTG for ordinary and the Quadratic Knapsack Problem -- the quadratic profits have to be taken into an account;
for each item $m' < m$ such that $p_{m m'} > 0$ we have to introduce a doubly-controlled addition of $p_{m m'}$, respectively: $C_{m m'}^{1}(\text{ADD}_{p_{m m'}})$.
These profit updates are all summarised as 
\begin{align*}
    U_{m}^{3} = C_{m}^{1}(C_{m - 1}^{1}(\text{ADD}_{p_{m m - 1}}) \cdots C_{1}^{1}(\text{ADD}_{p_{m 1}})\text{ADD}_{p_{m}}).
\end{align*}
The doubly-controlled additions may also be deferred after including/excluding all items and accounting for their linear profit, thus effectively singling out the original QTG proposed for the KP. 
The concept is summarized again in \hyperref[box:QTG-qkp]{Box 1}.

The just described QTG for the QKP now lays the foundation for applying quantum maximum finding (\textbf{QMaxSearch}) \cite{Durr1996AQuantumAlgorithmForFindingTheMinimum}, tailored to this very problem.
In a nutshell, \textbf{QMaxSearch} consists of the consecutive application of amplitude amplification (\textbf{QSearch}) \cite{Brassard2002QuantumAmplitudeAmplificationAndEstimation} where the oracle is given by coherent integer comparison of each solutions' total profit to the total profit of the incumbent such that solutions with higher profit are amplified.
The initial incumbent is obtained via a classical heuristic, e.g. a greedy packing method, and updated whenever the measurement after one round of amplitude amplification yields a better solution.
In our case, the QTG enters the amplitude amplification as state preparation circuit as well as part of the diffusion operator.

To benchmark the QTG-based search against its best classical competitors, we expand the high-level simulator introduced by Wilkening et~al.~\cite{Wilkening2023AQuantumAlgorithmForTheSolutionOfTheKnapsackProblem} to the QKP.
By following the resource estimation techniques from \cite{Wilkening2023AQuantumAlgorithmForTheSolutionOfTheKnapsackProblem}, we calculate the number of necessary quantum gates and cycles as well as the success probabilities of obtaining optimal solutions for various QKP instances.
The gate count is rather straight-forward as the QTG for the QKP effectively arises from the QTG for the KP plus additional doubly-controlled profit additions as shown in \autoref{figure:qknap_circuit}.
However, in the more cycle-efficient implementation, we directly incorporated the quadratic additions into the main layer unitaries $U_{m}$ instead of appending these operations.
This readily eliminates several QFTs and inverse QFTs between the individual profit additions.

% While the QTG state preparation aims to reduce the number of Grover iterations required to find incumbent solutions, improving the cycle count of the state preparation is equally important. 
In fact, we can reduce the cycle count of the quadratic profit additions even further by introducing additional qubits.
This is particularly relevant, as the main contributing factor for the runtime of the QTG-based search for the QKP is indeed summing up the objective function's $\mathcal{O}(n^{2})$ terms.
By introducing some ancilla qubits, the implementation cost of the QTG circuit can be lowered to be almost as cheap as for the KP:
We decompose the sum $p_{1} + p_{1,2} + \dots + p_{n, n - 2} + p_{n, n - 1}$ into pairwise additions.
Those are carried out by conditionally assigning every profit to its respective ancilla register and applying the addition circuit for every pairwise addition.
All these additions can be performed in parallel.
By using $\mathcal{O}(\log_{2}(n^2))$ pairwise addition layers, we can compute the objective function of every assignment in $\mathcal{O}(\log_{2}(n))$ time.
The same could be done for all the costs up to the break item $b = \min\left\{h: \sum_{m = 1}^{h} w_{m} > c\right\}$, since up to this item, every sub-assignment is feasible.
Further parallelization of the comparisons can be achieved by copying the capacity registers and performing the multi-controlled CNOT gates needed for the comparison in parallel.
Thus, by introducing $\mathcal{O}(\numbits{c}^{2})$ ancilla qubits in the worst case we obtain $\mathcal{O}(\log_{2} \numbits{c})$ cycles for the comparison.
The comparisons still have to be applied subsequently from item $b + 1$ to item $n$, still resulting in a $\mathcal{O}(n)$ runtime, but it does not show any quadratic circuit depth.

\begin{figure}[!ht]
    \begin{tcolorbox}[
        colback=white, 
        colframe=teal!40,
        coltitle=black, 
        title=\noindent\justifying{Box 2: QTG for Multidimensional Knapsack}]
    \begin{minipage}[t]{0.9\textwidth}
        \justifying\footnotesize
        \noindent\\\textbf{Input}\\
        Multidimensional Knapsack instance $\mdkp$.\\
        \\
        \textbf{1. Initialize:}\\
        Initialize the registers in the state $\ket{\bm{0}}^{1} \ket{\bm{c}}^{2} \ket{0}^{3}$.\\\\
        \textbf{2. Tree traversing:}\\
        For each item $m = 1, \ldots, n$:\\
        \textbf{(a)} Create a superposition: $U^{1}_{m} = \controlled^{2}_{\varphi_{m}}(R_{y, b}^{x_{m}})$\\\\
        \textbf{(b)} Update the capacities: $U^{2}_{m} = \controlled_{m}^{1}(\bigotimes_{i = 1}^{d}\text{SUB}_{w_{i m}}^{i})$\\\\
        \textbf{(c)} Update the total profit: $U^{3}_{m} = \controlled^{1}_{m}\left( \text{ADD}_{p_{m}} \right)$
    \end{minipage}
    \label{box:QTG-mkp}
    \end{tcolorbox}
\end{figure}

\section{\label{section:MultidimensionalKnapsack}Multidimensional Knapsack}

In the $0$-$1$-Multidimensional Knapsack Problem, we are again given a set of $n$ items.
As for the KP, these items carry (linear) integer profits $p_{m}$, but no quadratic/off-diagonal terms $p_{m m'}$ exist.
Instead, each item $m$ incurs $d$ integer weights that we denote as $w_{i m}$, $i = 1, \ldots, d$.
The MDKP asks for a subset of items of maximum cumulative profit, such that for each dimension $i$, the cumulative weights in dimension $i$ of all selected items do not exceed some capacity $c_{i}$.
Without loss of generality, we again assume that $\max_{m} w_{i m} \leq c_{i} < \sum_{m} w_{i m}$ for all dimensions $i$.
We address a given instance by $\mdkp$.
The corresponding integer linear program (ILP) can be formulated as follows.

\begin{equation*}
    \begin{aligned}
        \text { maximize } &\sum_{m = 1}^{n} p_{m} x_{m} & \\
        \text { subject to } & \sum_{m = 1}^{n} w_{i m} x_{m} \leq c_{i}, \quad i = 1, \ldots, d \\
        & x_{m} \in \{0, 1\}, \quad  m = 1, \ldots, n.
    \end{aligned}
\end{equation*}

For the paths $\bm{x} = x_{1} \cdots x_{n}$ and the associated profits we again utilize registers $\hil_{1}$ and $\hil_{3}$ of size $n$ and $\numbits{P}$, respectively, where $P$ is some upper bound on the optimal profit.
The only structural difference lies in the capacity register:
For each dimension $i$, we introduce a separate capacity register of respective size $\numbits{c_{i}}$, encoding the $i$-th remaining capacity $c_{i} - \sum_{m = 1}^{n} w_{i m} x_{m}$.
We summarize these $d$ capacity registers in a composite capacity register $\hil_{2}$.
The total number of qubits that the QTG method requires in the composite register $\hil = \hil_{1} \otimes \hil_{2} \otimes \hil_{3} \otimes \hil_{\ancilla}$ is
\begin{equation*}%\label{equation:TotalNumberOfQubitsMDKP} removed as it is not used anymore.
    n + \sum_{i = 1}^{d} \numbits{c_{i}} + \numbits{P} + \max(n, \sum_{i = 1}^{d} \numbits{c_{i}} + 1, \numbits{P}).
\end{equation*}
Ancilla qubits are introduced only to optimise circuit depth, as detailed in \cite{Wilkening2023AQuantumAlgorithmForTheSolutionOfTheKnapsackProblem}.

\subsection*{Methods}
We can construct the QTG $\QTG$ by following the same design rules as for KP and the QKP as a cascade of $n$ layer unitaries $U_{m} = U_{m}^{3} U_{m}^{2} U_{m}^{1}$.
$U_{m}$ includes/excludes item $m$ in superposition if and only if all $d$ remaining capacities remain non-negative upon subtracting the respective weight $w_{i m}$, and updates all remaining capacities as well as the total profit accordingly.
The branching unitary $U_{m}^{1}$ controls the creation of the exclusion/inclusion superposition on all $d$ comparisons ``$w_{i m} \leq c_{i}(\bm{x})$'' being valid.
Defining $\varphi_{m}$ to be the logical conjunction of all these $d$ comparisons, we denote the branching unitary $U_{m}^{1} = \controlled^{2}_{\varphi_{m}}(R_{y,b}^{x_m})$.
Subsequently, we update the $d$ remaining capacities by subtracting $w_{i m}$ from the respective capacity register's state, controlled on the $m$-th qubit, i.e.\ $U_{m}^{2} = \controlled_{m}^{1}(\bigotimes_{i = 1}^{d}\text{SUB}_{w_{i m}}^{i})$.
The profit updates are exactly the same as the linear updates for the QKP, that is $U_{m}^{3} = \controlled_{m}^{1}(\text{ADD}_{p_{m}})$.
The final state after applying all $n$ layer unitaries is the superposition of all states representing feasible solutions to the given instance of $\mdkp$, each with a certain amplitude depending on the chosen bias $b$.
The concept is summarized again in \hyperref[box:QTG-mkp]{Box 2}.

\begin{figure*}
    \centering
    \subfloat[Quadtratic Knapsack Problem\label{fig:qkp-comparison}]{%
    \includegraphics[width=.5\linewidth]{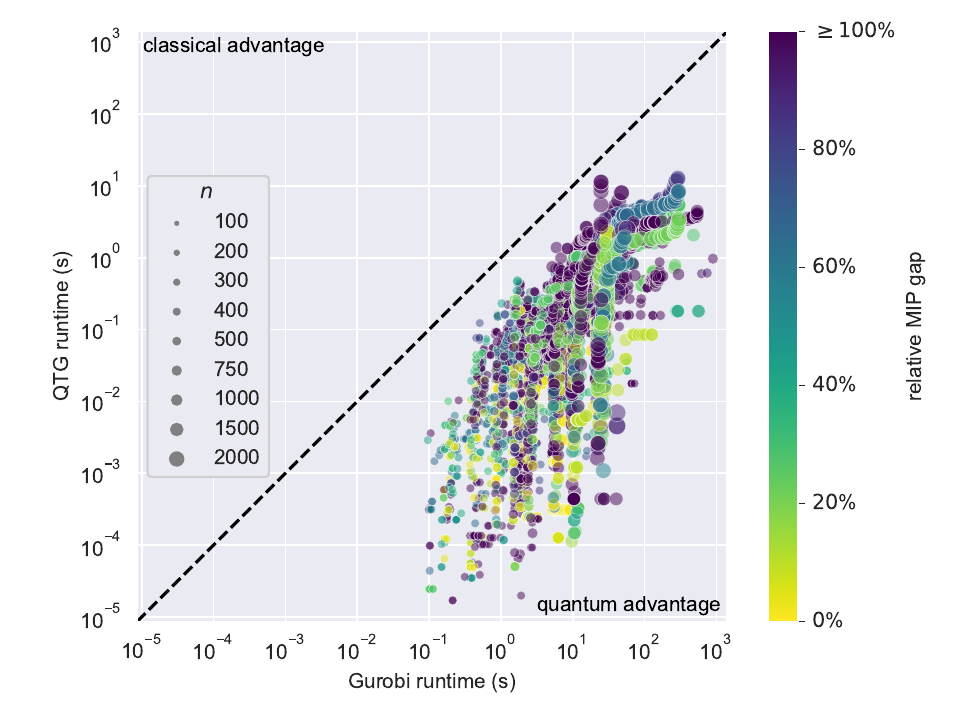}}
    \subfloat[Multidimensional Knapsack Problem\label{fig:mkp-comparison}]{%
    \includegraphics[width=.5\linewidth]{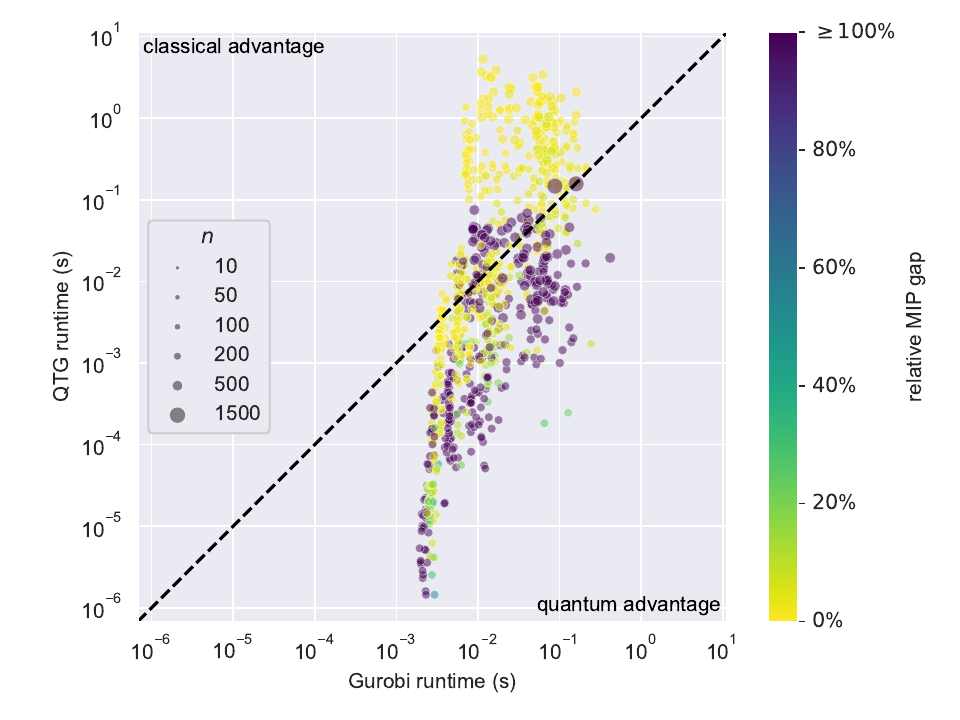}}
    \caption{QTG-based search vs Gurobi: performance analysis. We benchmark the algorithms on relevant benchmark instances with up to $2000$ variables for the QKP and $1500$ variables and $100$ constraints for the MDKP.
    Each data point represents the time required for Gurobi to obtain an incumbent solution and for the QTG-based search to find an equally good or better solution under benevolent assumptions for the quantum algorithm, see \autoref{subsection:AssumptionsOnTheQuantumProcessingUnit}.
    The markers' sizes and colors denote the instance's size and the solution's quality, respectively.
    Assuming a quantum cycle time limit of $1ns$ \cite{Chew2022UltrafastEnergyExchangeBetweenTwoSingleRydbergAtomsOnANanosecondTimescale}, we observe that for the given benchmark sets the QTG search has the potential to outperform the state-of-the-art solver Gurobi. 
    This is especially noticeable when dealing with problems that are more difficult to solve for classical algorithms (QKP).}
    \label{fig:qkp-mkp-comparison}
\end{figure*}

Similar to KP and QKP, the above QTG serves as the state preparation circuit in amplitude amplification (\textbf{QSearch}) \cite{Brassard2002QuantumAmplitudeAmplificationAndEstimation}, which is called as a subroutine in \textbf{QMaxSearch} \cite{Durr1996AQuantumAlgorithmForFindingTheMinimum}.
Starting with an initial solution and its profit as an initial threshold, we apply an integer comparison oracle to mark and amplify states with higher profits. 
The threshold is iteratively updated with the intermediate solutions' total profits.

\section{Results}

We demonstrate the capabilities of QTG on a diverse set of benchmark instances.
All experiments were carried out on a regular desktop workstation with an
AMD Ryzen 7 5800X ($8\times 3.8$ GHz) CPU and $128$ GB of RAM\@.
We slightly alter the benchmarking objective compared to \cite{Wilkening2023AQuantumAlgorithmForTheSolutionOfTheKnapsackProblem}.
We compare the solution finding capabilities of QTG with the Gurobi~\cite{Gurobi} solver.
To compare the classical and quantum sides, we use benevolent assumptions for the quantum algorithm explicitly stated in \autoref{subsection:AssumptionsOnTheQuantumProcessingUnit}.
We further expand the high-level simulator introduced by Wilkening~\textit{et~al.}~\cite{Wilkening2023AQuantumAlgorithmForTheSolutionOfTheKnapsackProblem}.
We chose the sampling-based method to assist our simulator.
By following the resource estimation techniques from \cite{Wilkening2023AQuantumAlgorithmForTheSolutionOfTheKnapsackProblem}, we calculate the number of necessary quantum gates and cycles as well as the success probabilities of obtaining optimal solutions.
To obtain a meaningful runtime for QTG, we assume a quantum cycle time of $1ns$~\cite{Chew2022UltrafastEnergyExchangeBetweenTwoSingleRydbergAtomsOnANanosecondTimescale}.

To monitor Gurobi's progress, we added a callback to save each incumbent solution that was found during solve, together with the time until the solution was found.
Similarly, whenever a new QTG solution is found, it is saved alongside the elapsed cycle count from which we can derive its runtime.
For the comparison in \autoref{fig:qkp-mkp-comparison}, we search for each incumbent solution from Gurobi an (earliest) QTG solution with equal or better objective value. If no such solution exists, the incumbent solution is ignored.
The scatterplot then compares Gurobi's runtime with the estimated QTG runtime to find an equally good solution.
The plots also highlight the quality of said solutions, i.e., the relative gap between the objective value $obj$ and the best known bound $bound$ ($|bound-obj| / |obj|$).

\paragraph{Quadratic Knapsack}
For the QKP, we chose the recent QKPLIB~\cite{qkplib,Jovanovic2023SolvingTheQuadraticKnapsackProblemUsingGRASP} benchmark set. It contains instances ranging from $100$ to $2000$ items. 
Each size has a total of $36$ instances, resulting in a total of $324$ instances. 
\autoref{fig:qkp-comparison} compares $3472$ incumbent solution from a subset of $238$ instances where QTG could find at least one solution with a better objective than Gurobi.
It shows potential quantum advantage for instances of all sizes and solution qualities that are close to the optimal solution ($<40\%$).
For solutions with a larger optimality gap, QTG finds better solutions several orders of magnitude faster than Gurobi, whereas for smaller optimality gaps the runtime of Gurobi and the estimated runtime are closer to each other.

\paragraph{Multidimensional Knapsack}
For the MDKP, we evaluated Gurobi and QTG for benchmark instances from \cite{drakeMKP} that consist of $335$ instances from the ORLIB~\cite{Chu1998AGeneticAlgorithmForTheMultidimensionalKnapsackProblem}, GK~\cite{Glover1996CriticalEventTabuSearchForMultidimensionlKnapsackProblems} and SAC-94 data sets. 
In \autoref{fig:mkp-comparison} we compare $944$ incumbent solutions from a subset of $332$ instances for which QTG could find at least one solution with a better objective than Gurobi.
\autoref{fig:mkp-comparison} shows that although quantum advantage is possible for some solutions with larger optimality gaps, QTG fails to deliver (close-to) optimal solutions faster than Gurobi for instances of all sizes.
It performs particularly poorly on high-quality solutions with a small optimality gap.

\paragraph*{Discussion}
Our results highlight both the potential and the limitations of QTG in comparison to classical solvers like Gurobi. 
The empirical evaluation demonstrates that, under benevolent assumptions, QTG exhibits promising solution-finding capabilities, especially in problem instances where optimality gaps remain moderate. 
For the QKP benchmark, we observe a quantum advantage in many cases, with QTG identifying high-quality solutions significantly faster than Gurobi, particularly when the required optimality gap is below 40\%.

However, our evaluation also underscores the limits of QTG when facing powerful classical algorithms for problems it is well-suited for such as the MDKP. 
For these instances, even under favorable assumptions, QTG struggles to match Gurobi's ability to find high-quality solutions efficiently. 
The results indicate that as the solution quality approaches optimality, the relative performance gap between QTG and Gurobi increases, with Gurobi often maintaining an advantage.

\section{\label{section:Conclusion}Conclusion}

We extended the scope of the Quantum Tree Generator to the $0$-$1$ Quadratic Knapsack (QKP) and the $0$-$1$ Multidimensional Knapsack Problem (MDKP).
That is, for both problem classes, we constructed a polynomial-depth quantum circuit generating a (generally non-uniform) superposition of all feasible solutions with their respective profits stored in an ancilla register.
As in its original proposal, using the QTG as state preparation routine within quantum maximum finding yields a quantum optimization algorithm for the respective problem: QTG-based search.
We empirically evaluated the capabilities of QTG-based search for both problem classes against a state-of-the-art classical solver on realistic instances.
Our findings are twofold:
First, we found, assuming fault-tolerant, fully connected qubits as well as fully parallelizable quantum cycles with $1ns$ clock speed, that QTG-based search outperforms its classical competitor across all considered instances of QKP.
Second, under the same quantum hardware assumptions, we observed that for the MDKP there are instance-specific factors that can shift the advantage to either the classical or quantum side.
Since the QTG circuits for both problems are very similar, it is probably the difficulty the classical algorithm has with the non-linear objective function of the QKP tipping the scale in favor of the QTG-based search in the first, but not in the second case.
Therefore, promising candidates for more significant quantum speed-ups might be problems where classical algorithms fail to produce high-quality solutions, such as problems with non-convex objective functions.
While quantum computing holds promise for combinatorial optimization, our study indicates that achieving a definitive advantage over classical algorithms remains a challenge that will require further theoretical and technological advancements.

\section*{Acknowledgment}
This work was supported by the DFG through SFB 1227 (DQ-mat) and Germany’s Excellence Strategy via EXC-2123 QuantumFrontiers 390837967 and PhoenixD, as well as the Quantum Valley Lower
Saxony, the BMBF project ATIQ, the BMBF project QuBRA, the BMWi project ProvideQ, and the ERC and the DFG via the project ResourceQ.

\section*{Author contributions}
This project was conceived of, and initiated in, discussions of S.W. and T.J.O. 
The quantum implementation along with the resource estimation procedures were developed by S.W, A.I.L, L.B., M.F., and R.K.. 
The classical expertise was provided by S.F. and M.P. All authors contributed to writing the paper.

\IEEEtriggeratref{36}
\bibliographystyle{IEEEtran}
\bibliography{main.bib}

\end{document}